\documentclass[10pt,conference]{IEEEtran}
\IEEEoverridecommandlockouts
\usepackage{cite}
\usepackage{balance}
\usepackage[cmex10]{amsmath}
\usepackage{amssymb,amsfonts}
\usepackage{bm}        % bold math symbols
\usepackage{bbm}
\usepackage{amsthm}    % theorem environments

\usepackage{algorithmic}
\usepackage{algorithm}

\usepackage{graphicx}

\usepackage{multirow,booktabs,makecell}

\usepackage{enumitem}

\usepackage{xcolor}
\usepackage{soul}

\usepackage[american]{babel}
\usepackage[T1]{fontenc}
\usepackage[utf8]{inputenc}

\usepackage{siunitx}

\usepackage{textcomp}

\usepackage{microtype}
\def\BibTeX{{\rm B\kern-.05em{\sc i\kern-.025em b}\kern-.08em
    T\kern-.1667em\lower.7ex\hbox{E}\kern-.125emX}}

\theoremstyle{plain}

\theoremstyle{definition}

\theoremstyle{remark}

\newcommand*{\herm}{^{\mathsf{H}}}
\newcommand*{\transp}{^{\mathsf{T}}}

\newcommand{\e}{\mathrm{e}}

\usepackage{fancyhdr}
\IEEEoverridecommandlockouts
\fancypagestyle{myfancy}{
	\fancyhf{} % clear all header and footer fields
	\fancyhead[C]{\textit{Accepted to the 32nd International Conference on Telecommunications (ICT 2026), Thessaloniki, Greece}} % center header
	\fancyfoot[C]{\footnotesize © 2026 IEEE. Personal use of this material is permitted. 
		Permission from IEEE must be obtained for all other uses.}

}

\begin{document}

\title{Waveform Index Modulation for Backscatter Communications in RIS-Based MIMO Radars}

\author{
\IEEEauthorblockN{
Mehri~Nikzad\IEEEauthorrefmark{1}\IEEEauthorrefmark{2}, 
Luca~Venturino\IEEEauthorrefmark{1}\IEEEauthorrefmark{2}\IEEEauthorrefmark{3}, 
Emanuele~Grossi\IEEEauthorrefmark{1}\IEEEauthorrefmark{2}\IEEEauthorrefmark{3}, 
Xiaodong~Wang\IEEEauthorrefmark{4}
}

\IEEEauthorblockA{\IEEEauthorrefmark{1}
\textit{University of Cassino and Southern Lazio, 03043 Cassino, Italy}
}

\IEEEauthorblockA{\IEEEauthorrefmark{2}
	\textit{National Inter-University Consortium for Telecommunications (CNIT), 43124 Parma, Italy}
}

\IEEEauthorblockA{\IEEEauthorrefmark{3}
	\textit{People Oriented Smart Technology Lab (POSTLab), European University of Technology (EUt+)}
}

\IEEEauthorblockA{\IEEEauthorrefmark{4}
\textit{Columbia University, New York, NY 10027, USA}
}

\IEEEauthorblockA{
E-mails: mehri.nikzad@unicas.it, l.venturino@unicas.it, e.grossi@unicas.it, xw2008@columbia.edu
}

\thanks{
The work of M.~Nikzad was supported by the European Union through the project ISLANDS (Grant agreement n. 101120544).  The work of L.~Venturino and X.~Wang was supported by the U.S. National Science Foundation (NSF) under Grant ECCS-2335765.
The work of E.~Grossi was supported by the project ``FLARE'' funded by the European Union under the Italian National Recovery and Resilience Plan of NextGenerationEU (program ``RESTART'').
}
}

\maketitle
\thispagestyle{myfancy}

%%%
\bstctlcite{BSTcontrol}	 % control bibliography style settings

\begin{abstract}
This paper studies a symbiotic system in which a reconfigurable intelligent surface (RIS) assists a radar transmitter while conveying information to a reader via backscattering. The RIS is partitioned into subarrays that redirect the radar signal toward the angular sector under inspection and superimpose a slow-time modulation using orthogonal phase codes, thereby implementing MIMO radar functionalities. Communication is achieved by encoding information in the selection of an unordered subset of orthogonal codewords, without altering the RIS transmit beampattern. At the reader, the proposed index modulation scheme enables low-complexity detection without requiring channel state information. Numerical results demonstrate the effectiveness of the proposed backscatter communication approach.
\end{abstract}

\begin{IEEEkeywords}
MIMO radar, reconfigurable intelligent surface (RIS), backscatter communication, index modulation, symbiotic radar-communication systems.
\end{IEEEkeywords}

\section{Introduction}

Ambient backscatter communication (ABC) is a promising technology for enabling low-power and battery-free communications~\cite{Huynh2018}, wherein a backscatter device (tag) conveys information to a destination (reader) by modulating and reflecting  signals emitted by existing radio-frequency sources~\cite{Stockman-1948,nikitin2008antennas,Liu13ambientbackscatter}. In this context, reconfigurable intelligent surfaces (RISs) provide a natural implementation of backscatter modulators~\cite{BackCom-RIS-2022}. Indeed, an RIS comprises a large number of programmable elements whose reflection coefficients can be jointly controlled, enabling space-time modulation of the impinging waveform~\cite{Liu_RIS_ComSurTut2021,Mizmizi_STcodedRIS_JSAC2024,Chen_STcodedRIS_Nature2025}.

RISs have also been extensively studied to enhance legacy wireless systems for communication~\cite{Basar_Com-RIS_Access2019,Yu_RIS_ISAC_TCOM2023}, sensing~\cite{Aubry-2021,Buzzi_RIS_SP2022}, and integrated sensing and communication (ISAC)~\cite{Chepuri_RIS-ISAC_SPM2023,Taremizadeh_STAR-RIS_OJCS2025}. When integrated into the transmitter, they provide a large effective aperture at low cost, while, when deployed in the environment, they enable non-line-of-sight operation via programmable signal redirection. These capabilities naturally extend the ABC paradigm, as the same reflection mechanism used for information embedding can also shape propagation and support sensing~\cite{Wang-2023}.
This synergy is particularly appealing in scenarios involving mobile agents, such as vehicular systems and robotic platforms, where many low-cost devices must be connected with minimal energy consumption and signaling overhead~\cite{Deng_RIS-vehicular_TIV2025}. Representative applications include smart transportation systems and indoor environments such as factories and warehouses. In such settings, passive tags embedded in the environment (e.g., road elements, machinery, or safety devices) can convey information through ABC while supporting the underlying wireless infrastructure, effectively realizing a symbiotic system~\cite{Liang_Symbiotic-Radio_TCCN2020,Ren_Symbiotic-Radio_ITJ2023}.

Motivated by these observations, we investigate a symbiotic system in which an RIS-based device operates both as a passive relay to assist a legacy radar transmitter (source)~\cite{Aubry-2021,Buzzi_RIS_SP2022} and as a passive tag to implement radar-enabled backscatter communication~\cite{Venturino_RadBackCom_TWC2023,Venturino_RadBackCom_TWC2024}. The RIS is partitioned into subarrays that implement MIMO radar functionalities through orthogonal slow-time modulation~\cite{LiStoica2007,Book_StoicaLi_2009}. We propose a waveform index modulation scheme in which information is encoded in the unordered selection of subsets of orthogonal slow-time codewords assigned to the subarrays. By decoupling communication from spatial beamforming, the proposed design preserves the RIS transmit beampattern. At the reader, this structure enables a low-complexity noncoherent detection rule based on a bank of matched filters, avoiding channel state information (CSI) acquisition. The transmitted message is recovered by identifying the set of active codewords through energy detection. The error probability is analytically derived, and numerical results illustrate the achievable transmission rates and error performance as functions of the codeword length.

The remainder of this paper is organized as follows.\footnote{In the following, $\mathbb{C}$ denotes the set of complex numbers, and $\mathbb{U} =\{z \in \mathbb{C} : |z| = 1\}$ denotes the set of unit-modulus complex numbers. Column vectors and matrices are denoted by lowercase and uppercase boldface letters, respectively. The symbols $(\,\cdot\,)\transp$ and $(\,\cdot\,)\herm$ denote transpose and conjugate-transpose, respectively. The vector $\bm{1}_{M}$ denotes the $M$-dimensional all-ones column vector. The entry in the $i$-th row and $j$-th column of $\bm{A}$ is denoted by $[\bm{A}]_{i,j}$. The operators $\star$, $\odot$, $\|\cdot\|$, and $\mathrm{E}[\cdot]$  denote the convolution, Hadamard product, Euclidean norm, and statistical expectation, respectively.} Sec.~\ref{SEC_System_Description} describes the system model. Sec.~\ref{SEC_System_Design} presents the proposed modulation scheme at the tag and the corresponding detector at the reader. Sec.~\ref{SEC_Numerical_Analysis} presents the numerical analysis. Finally, concluding remarks are given in Sec.~\ref{SEC_Conclusions}.

\section{System Description}\label{SEC_System_Description}

\begin{figure}[!t]
	\centerline{\includegraphics[width=0.8\columnwidth]{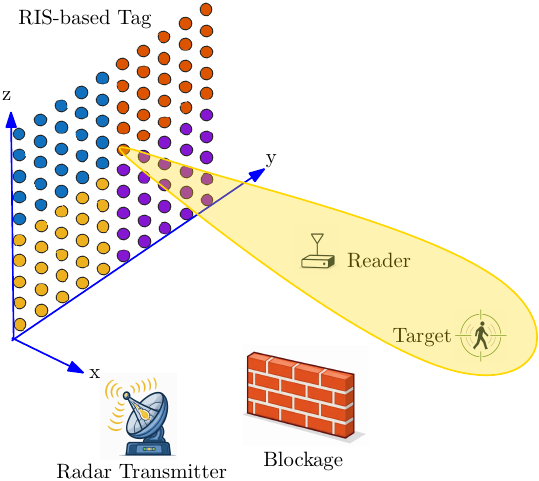}}
    \vspace{-0.1cm}
	\caption{Considered system architecture.}
	\label{Fig_01_proposed_system}
\end{figure}
We consider the symbiotic system in Fig.~\ref{Fig_01_proposed_system}. The RIS-based tag assists a single-antenna radar transmitter (hereafter referred to as the \emph{source}) in illuminating an angular sector that would otherwise be unreachable~\cite{Aubry-2021,Buzzi_RIS_SP2022}. The $M_{\rm RIS}=N M$ elements of the RIS are partitioned into $N$ subarrays, each comprising $M$ elements and applying a slow-time modulation to the reflected signal. By employing orthogonal modulation waveforms, this configuration effectively realizes an RIS-aided MIMO radar. At the same time, the collaborative tag exploits the source signal to establish a backscatter communication with a reader, while preserving sensing functionality.

\subsection{Transmitted signal}
The baseband probing signal emitted by the source is a periodic pulse train with pulse repetition interval (PRI) $T_{\rm PRI}$. The transmitted pulse has waveform $a(t)$, power $\mathcal{P}$, bandwidth $W$, and duration $T = G/W \ll T_{\rm PRI}$, where the integer $G \geq 1$ determines the time–bandwidth product (also referred to as the processing gain).
Tag transmissions toward the reader are organized into frames, each consisting of $L$ PRIs. For illustration, we consider the frame interval $[0,LT_{\rm{PRI}}]$.

Assuming that the RIS aperture is much smaller than $c/W$ (narrowband assumption~\cite{Book_Stutzman_2012}), where $c$ denotes the speed of light, and that there is a dominant line-of-sight path between the source and the tag, the source--tag (ST) channel is modeled as $\bm{\gamma}_{\rm ST}(t)=\bm{\gamma}_{\rm ST}\,\delta(t) \in \mathbb{C}^{M_{\rm RIS}}$, where the vector 
\begin{equation}
\bm{\gamma}_{\rm ST} = [\gamma_{\rm ST,1}; \ldots; \gamma_{\rm ST,M_{\rm RIS}}]
\end{equation}
accounts for the propagation effects and the antenna gain of the transmitter and the RIS elements.\footnote{The propagation delay between the source and the tag can be absorbed into the link from the tag to the reader and is omitted for brevity.}
We assume that the ST channel varies over time scales much longer than the frame duration, thus enabling its estimation. Such channel knowledge is required to steer the reflected signal toward a desired angular sector~\cite{Aubry-2021,Buzzi_RIS_SP2022,Yu_RIS_ISAC_TCOM2023,Delia_RIS_TVT2025}. In contrast, the ST channel is unavailable at the reader to limit signaling overhead in the implementation of backscatter communication~\cite{Venturino_RadBackCom_TWC2023,Venturino_RadBackCom_TWC2024}.

The signal reflected by the $m$-th RIS element is
\begin{equation}
	\sum_{\ell=0}^{L-1} x_{\ell,m}\,\gamma_{{\rm ST},m}\, a\big(t - \ell T_{\rm PRI}\big),
	\label{Eq_phi_m}
\end{equation}
where $x_{\ell,m}$ denotes the unit-modulus response during the $\ell$-th PRI and $\gamma_{{\rm ST},m}(t)$ is the $m$-entry of $\bm{\gamma}_{\rm ST}(t)$. We denote by $\bm{X}\in\mathbb{U}^{L\times M_{\rm RIS}}$
the \emph{code matrix} describing the RIS response over the frame interval, with $[\bm{X}]_{\ell,m}=x_{\ell,m}$.

\subsection{Received signal at the reader}

Assume that the source--reader (SR) and tag--reader (TR) channels remain constant over a frame interval, and denote by $\gamma_{\rm SR}(t)\in\mathbb{C}$ and $\bm{\gamma}_{\rm TR}(t) = [\gamma_{{\rm TR},1}(t); \cdots; \gamma_{{\rm TR},M_{\rm RIS}}(t)] \in \mathbb{C}^{M_{\rm RIS}}$ their baseband impulse responses, with support $[\Delta_{\rm SR}^{\min}, \Delta_{\rm SR}^{\max}]$ and $[\Delta_{\rm TR}^{\min}, \Delta_{\rm TR}^{\max}]$, respectively. Also, define
\begin{subequations}
	\begin{align}
		\alpha_{{\rm STR},m}(t) &= a(t)\star \gamma_{{\rm ST},m}(t) \star \gamma_{{\rm TR},m}(t), \label{alpha_m}\\
		i_{\rm SR}(t) &= a(t) \star \gamma_{\rm SR}(t),
	\end{align}
\end{subequations}
which correspond to the filtered source pulse received at the reader via the indirect path associated with the $m$-th RIS element, and via the direct SR path, respectively. The received baseband signal at the reader is then
\begin{align}
	y_{\rm R}(t) &= \underbrace{\sum_{\ell=0}^{L-1} \sum_{m=1}^{ M_{\rm RIS}} x_{\ell,m}\,\alpha_{{\rm STR}, m}(t-\ell T_{\rm PRI})}_{\text{signal of interest}} \notag \\
	&\quad + \underbrace{\sum_{\ell=0}^{L-1} i_{\rm SR}(t-\ell T_{\rm PRI})}_{\text{source interference}} + \omega_{\rm R}(t),
	\label{reader_rx_signal}
\end{align}
where the first term contains the tag message, the second term represents the interference from the source, and $\omega_{\rm R}(t)$ denotes additive noise. Under the mild conditions
\begin{subequations}\label{cond_reader}
	\begin{align}
		T_{\rm PRI} &> T + \Delta_{\rm TR}^{\max}, \label{cond_reader_1}\\
		T_{\rm PRI} &> T + \Delta_{\rm SR}^{\max}, \label{cond_reader_2}
	\end{align}
\end{subequations}
signals associated with different PRIs do not overlap in~\eqref{reader_rx_signal}.  
This typically occurs in short-range scenarios, where the source, tag, and reader are in close proximity so that the propagation delays of the dominant paths are negligible compared to $T_{\rm PRI}-T$. The conditions are also satisfied when $T_{\rm PRI}$, controlled by the source, is designed to exceed the maximum propagation delay over the coverage region, as commonly done in radar systems to ensure an unambiguous range.

The received signal $y_{\rm R}(t)$ is sampled at time instants
$\Delta_{\rm TR}^{\min} + \ell T_{\rm PRI} + k/W$, for $\ell = 0, \ldots, L-1$ and $k = 0, \ldots, K_{\rm R}-1$,
where $W$ is the Nyquist sampling rate and
\begin{equation}
	K_{\rm R} = \left\lceil \big(T + \Delta_{\rm TR}^{\max} - \Delta_{\rm TR}^{\min}\big) W \right\rceil
\end{equation}
is the number of samples per PRI. These samples are arranged into the matrix $\bm{Y}_{\rm R} \in \mathbb{C}^{L \times K_{\rm R}}$ as
\begin{equation}
	[\bm{Y}_{\rm R}]_{\ell+1,k+1} = y_{\rm R}\big(\Delta_{\rm TR}^{\min} + \ell T_{\rm PRI} + k/W\big).
\end{equation}
\textcolor{blue}{}
Under the conditions in~\eqref{cond_reader}, we have~\cite{Venturino_RadBackCom_TWC2023}
\begin{equation}
	\bm{Y}_{\rm R} = \bm{X} \bm{A}_{\rm STR}\herm + \bm{1}_{L} \bm{i}_{\rm SR}\herm + \bm{\Omega}_{\rm R},
	\label{reader_discrete_signal}
\end{equation}
where $\bm{A}_{\rm STR} = [\bm{\alpha}_{{\rm STR},1}, \ldots, \bm{\alpha}_{{\rm STR},M_{\rm RIS}}] \in \mathbb{C}^{K_{\rm R} \times M_{\rm RIS}}$, $\bm{\alpha}_{{\rm STR},m} \in \mathbb{C}^{K_{\rm R}}$ and $\bm{i}_{\rm SR} \in \mathbb{C}^{K_{\rm R}}$ collect the samples of $\alpha_{{\rm STR},m}(t)$ and $i_{\rm SR}(t)$ at time instants $\Delta_{\rm TR}^{\min} + k/W$, for $k = 0, \ldots, K_{\rm R}-1$, respectively, and $\bm{\Omega}_{\rm R}$ is the noise matrix whose entries are modeled as independent circularly symmetric complex Gaussian random variables with variance $\sigma_{{\rm R},\omega}^{2}$.

\section{System Design}\label{SEC_System_Design}

This section describes the proposed encoding strategy at the tag and the corresponding decoding rule at the reader.

\subsection{Encoding rule at the tag}

Let $\mathcal{M}=\{1,\ldots,M_{\rm RIS}\}$ denote the set of RIS elements, and let $\mathcal{M}_n=\{\mu_{n,1},\ldots,\mu_{n,M}\}$ denote the subset associated with the $n$-th subarray. Moreover, let $\bm{c}_n\in\mathbb{U}^{L}$ denote the slow-time phase-code assigned to the $n$-th subarray, and let $\bm{b}_n\in\mathbb{C}^{M}$ denote the corresponding spatial beamformer, designed according to the sensing illumination requirements. Then, the RIS space-time code matrix can be written as
\begin{equation}
	\bm{X}=\sum_{n=1}^{N}\bm{c}_n\big(\bm{P}_n\bm{b}_n\big)\herm,
	\label{proposed_enc_rule}
\end{equation}
where $\bm{P}_n\in\{0,1\}^{M_{\rm RIS}\times M}$ embeds $\bm{b}_n$ into the full RIS aperture according to the subarray structure, i.e.,
\begin{equation}
	[\bm{P}_{n}]_{m,j}=
	\begin{cases}
		1, & \text{if } m=\mu_{n,j},\\
		0, & \text{otherwise},
	\end{cases}
\end{equation}
for $m=1,\ldots,M_{\rm RIS}$ and $j=1,\ldots,M$.

Backscatter communication is implemented in the slow-time domain. To this end, consider a codebook
\begin{equation}
	\mathcal{U}=\{\bm{u}_1,\ldots,\bm{u}_{L-1}\},
\end{equation}
whose codewords are unit-modulus, orthogonal to $\bm{1}_L$, and mutually orthogonal. 
At each frame, the tag selects an \emph{unordered} set of $N$ distinct codewords from $\mathcal{U}$ and assigns one codeword to each subarray. 
The transmitted message is encoded solely in the \emph{selected codewords}, since the reader cannot uniquely associate them with specific subarrays due to the lack of knowledge of the RIS partitioning and spatial responses, and the lack of CSI.
The transmission rate is
\begin{equation}
	\mathcal{R}=\frac{1}{L}\log_2\binom{L-1}{N}\quad\text{[bit/PRI]}.
\end{equation}

The orthogonality of the adopted codewords to $\bm{1}_L$ is required to ensure that the signal of interest can be separated from the source interference. Moreover, mutual orthogonality among the codewords is required to support MIMO radar functionalities. In particular, the proposed encoding strategy preserves the RIS beampattern. To make this explicit, let $\bm{\psi}(\bm{\theta}) \in \mathbb{U}^{M_{\rm RIS}}$ denote the RIS steering vector toward direction $\bm{\theta} = [\theta^{\rm az}; \theta^{\rm el}]$, where $\theta^{\rm az}$ and $\theta^{\rm el}$ are the azimuth and elevation angles, respectively. Then, the energy radiated by the RIS toward direction $\bm{\theta}$ over one frame interval is proportional to 
\begin{align}
	\mathrm{B}(\bm\theta)
	&=	
	\left\|
	\bm X (\bm{\gamma}_{\rm ST} \odot \bm{\psi}(\bm{\theta}))
	\right\|^2\notag\\
	&=
	L\sum_{n=1}^{N}
	\left|
	(\bm{\gamma}_{\rm ST} \odot \bm{\psi}(\bm\theta))\herm\bm P_n \bm b_n
	\right|^2, \label{beam_pattern}
\end{align}
where the second equality follows from the orthogonality of the selected codewords. Hence, the RIS beampattern is independent of the selected codewords and is controlled via $\{\bm b_n\}_{n=1}^{N}$. 

\subsection{Data detection at the reader}\label{SEC: Data decoder at the reader}
Substituting~\eqref{proposed_enc_rule} into~\eqref{reader_discrete_signal}, the received signal is written as
\begin{equation}
	\bm Y_{\rm R} = \sum_{n=1}^{N} \bm c_n \bm \beta_{{\rm STR},n}\herm + \bm 1_L \bm i_{\rm SR}\herm + \bm \Omega_{\rm R},
\end{equation}
where $\bm \beta_{{\rm STR},n}= \bm A_{\rm STR} \bm P_n \bm b_n $.
The vectors $\{\bm \beta_{\rm STR} \}_{n=1}^{N}$ and $\bm i_{\rm SR}$ are unknown, since they depend on tag parameters and propagation channels unavailable at the reader. Hence, data detection must be performed in a noncoherent fashion~\cite{Qian-2017a}. 

\begin{figure}[!t]
	\centerline{\includegraphics[width=\columnwidth]{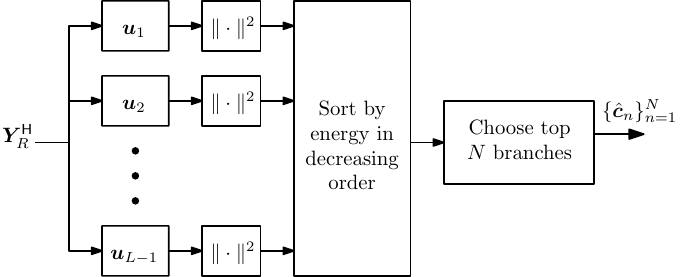}}
	\caption{Proposed detector at the reader.}
	\label{Fig_03_data_decoder_1}
\end{figure}

For any $\bm u \in \mathcal U$, consider the matched-filter output
\begin{equation}
	\bm Y_{\rm R}\herm  \bm u =
	\begin{cases}
		L \bm \beta_{{\rm STR},n}  + \bm \Omega_{\rm R}\herm \bm u, & \text{if } \bm{u}=\bm{c}_n,\\
		\bm \Omega_{\rm R}\herm \bm u, & \text{otherwise}.
	\end{cases}
\end{equation}
Note that the radar interference is removed after matched filtering, since we impose $\bm 1_L\herm \bm u = 0$ for any $\bm u \in \mathcal{U}$. Due to the orthogonality of the vectors $\{\bm{u}_{\ell}\}_{\ell=1}^{L-1}$, the output energy $\|\bm Y_{\rm R}\herm \bm u\|^2$ is large when $\bm u$ is a transmitted codeword and small otherwise. Accordingly, the reader computes $\|\bm Y_{\rm R}\herm \bm u\|^2$ for all $\bm u \in \mathcal U$ and detects the $N$ codewords corresponding to the largest energies, as shown in Fig.~\ref{Fig_03_data_decoder_1}. The computational complexity of this detector is $\mathcal{O}(L^2 K_{\rm R})$. 

The error probability, defined as the probability that the unordered set of detected codewords does not match the transmitted one, is expressed as
\begin{align}
	P_{\rm e} &= 1- \mathrm{E}\bigg[\int_{0}^{+\infty} 
	\prod_{n=1}^{N} Q_{K_{\rm R}}\!\left( \sqrt{2\,\frac{L\,\|\bm \beta_{{\rm STR},n} \|^{2}}{\sigma_{{\rm R},\omega}^{2}}}, \sqrt{2x} \right)\nonumber\\
	&\quad \times (L-N-1)\, f(x)\,F(x)^{L-N-2}\, dx\bigg],
	\label{Pc_known_rate}
\end{align}
where the expectation is taken over $\{\bm \beta_{{\rm STR},n}\}_{n=1}^{N}$, $Q_{K_{\rm R}}(\cdot,\cdot)$ denotes the generalized Marcum $Q$-function of order $K_{\rm R}$, and $f(x)$ and $F(x)$ are the probability density function (PDF) and the cumulative distribution function (CDF) of a Gamma-distributed random variable with shape parameter $K_{\rm R}$ and unit scale~\cite{Book_Papoulis_2002}. The derivation of~\eqref{Pc_known_rate}  is outlined next.

\subsubsection{Derivation of the error probability}

With reference to the $\ell$-th branch of the block diagram in Fig.~\ref{Fig_03_data_decoder_1}, define the statistic
\begin{equation}
T_{{\rm R},\ell} = \frac{\left\| \bm{u}_{\ell}\herm \mathbf{Y}_{\rm R} \right\|^{2}}{L \sigma^{2}_{{\rm R},\omega}}.
\label{statistics}
\end{equation}
Without loss of generality, assume $\bm{c}_n=\bm{u}_{n}$ for $n=1,\ldots,N$. For $\ell=1,\ldots,N$ and conditioned on $\bm \beta_{{\rm STR},\ell}$, the random variable $2T_{{\rm R},\ell}$ follows a noncentral chi-square distribution with $2K_{\rm R}$ degrees of freedom and noncentrality parameter $2 L\,\|\bm \beta_{{\rm STR},\ell}\|^{2} / \sigma_{{\rm R},\omega}^{2}$. For $\ell=N+1, \ldots,L-1$, $T_{{\rm R},\ell}$ follows a Gamma distribution with shape parameter $K_{\rm R}$ and unit scale. Due to the orthogonality of the vectors $\{\bm{u}_{\ell}\}_{\ell=1}^{L-1}$, the statistics $\{T_{{\rm R},\ell}\}_{\ell=1}^{L-1}$ are independent. To proceed, note that the conditional complementary CDF of $T_{{\rm R},\ell}$ is
\begin{equation}
\Pr\left\{{T_{{\rm R},\ell}}>x\,|\, \bm \beta_{{\rm STR},\ell} \right\} =
Q_{K_{\rm R}}\!\left(\sqrt{2\,\frac{L\,\|\bm \beta_{{\rm STR},\ell}\|^{2}}{\sigma_{{\rm R},\omega}^{2}}},\,\sqrt{2x}\right),
\end{equation}
for $\ell=1,\ldots,N$. Also, define the maximum among the last $L-N-1$ statistics in~\eqref{statistics} as
\begin{equation}
T_{{\rm R},\max} = \max_{\ell \in \{N+1, \ldots, L-1\}} T_{{\rm R},\ell},
\end{equation}
whose PDF is $(L-N-1)F^{\,L-N-2}(x)f(x)$.
A correct decision occurs when all statistics $T_{{\rm R},1},\ldots,T_{{\rm R},N}$ exceed $T_{{\rm R},\max}$. Hence, the error probability is expressed as in~\eqref{Pc_known_rate}.

\section{Numerical Analysis}\label{SEC_Numerical_Analysis}
We consider a system operating at a carrier frequency of $\qty{24}{\giga\hertz}$ with PRI $T_{\rm PRI}=\qty{3}{\micro\second}$. The waveform $a(t)$ is a phase-coded pulse generated using an $m$-sequence of length $G=15$, with bandwidth $W=\qty{50}{\mega\hertz}$ and duration $T=\qty{0.3}{\micro\second}$. The tag is equipped with a square RIS comprising $M_{\rm RIS}=225$ elements with half-wavelength spacing, partitioned into $N=9$ square subarrays. 
The ST channel is modeled as $\bm{\gamma}_{\rm ST} = \sigma_{\rm ST} e^{j\phi_{\rm ST}}\bm{\psi}(\bm{\theta}_{\rm ST})$, where $\sigma_{\rm ST}$ accounts for free-space attenuation, $\phi_{\rm ST}$ is uniformly distributed over $[0,2\pi)$, and $\bm{\theta}_{\rm ST} = [-45^\circ;0^\circ]$. The spatial beamformer of each RIS subarray is designed to maximize the radiated energy toward $\bar{\bm{\theta}} = [45^\circ;0^\circ]$. Accordingly, from~\eqref{beam_pattern}, we set
$\bm{b}_{n}=\bm P_n\herm \big(\bm{\psi}(\bm{\theta}_{\rm ST}) \odot \bm{\psi}(\bar{\bm{\theta}})\big)$, for $n=1,\ldots,N$.
Finally, the TR channel is modeled as
\begin{equation}
	\bm{\gamma}_{\rm TR}(t)=
	\sum_{q=0}^{Q_{\rm TR}-1} \gamma_{{\rm TR},q}\bm{\psi}(\bm{\theta}_{{\rm TR},q})\delta\big(t-\Delta_{\rm TR}^{\min}-\tau_{{\rm TR},q}\big),
\end{equation}
where $Q_{\rm TR}$ is the number of channel taps, and $\gamma_{{\rm TR},q}$, $\bm{\theta}_{{\rm TR},q}$, and $\tau_{{\rm TR},q}$ denote the complex amplitude, departure angle, and the delay offset of the $q$-th, respectively. 
The channel taps are independently generated. The complex amplitude is modeled as the sum of a
specular and a diffuse components~\cite{Shnidman-1999}, i.e.,
\begin{equation}
	\gamma_{{\rm TR},q}= \sigma_{\rm TR}\left(\sqrt{\frac{\kappa_{\rm TR}}{1+\kappa_{\rm TR}}}\e^{j\phi_{{\rm TR},q}}
	+\sqrt{\frac{1}{1+\kappa_{\rm TR}}} g_{{\rm TR},q}\right),
\end{equation}
where $\sigma_{\rm TR}^2$ denotes the average power, $\kappa_{\rm TR}$ is the power ratio between the specular and diffuse components, $\phi_{{\rm TR},q}$ is uniformly distributed in $[0,2\pi)$, and $g_{{\rm TR},q}$ is a complex circularly-symmetric Gaussian random variable with unit variance. 
The departure angle is uniformly distributed in
$[33^\circ,57^\circ]\times[-12^\circ,12^\circ]$. The delay offset is uniformly distributed in $[0,\Delta_{\rm TR}^{\max}-\Delta_{\rm TR}^{\min}]$. In the simulations, we set $\Delta_{\rm TR}^{\max}-\Delta_{\rm TR}^{\min}=15/W$, yielding $K_{\rm R}=30$ samples per PRI, $Q_{\rm TR}=3$, and $\kappa_{\rm TR}=10$~dB. 

\begin{figure}[!t]
	\centerline{\includegraphics[width=0.9\columnwidth]{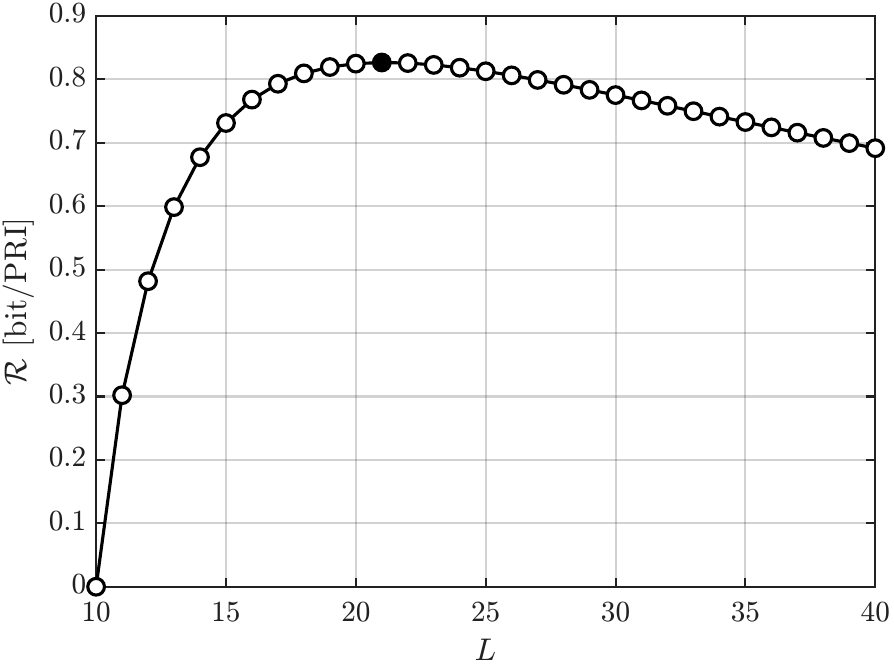}}
    \vspace{-0.2cm}
	\caption{Transmission rate $\mathcal{R}$ versus codeword length $L$.}
	\label{Fig_03_Rate_vs_L}
\end{figure}
\begin{figure}[!t]
	\centerline{\includegraphics[width=0.9\columnwidth]{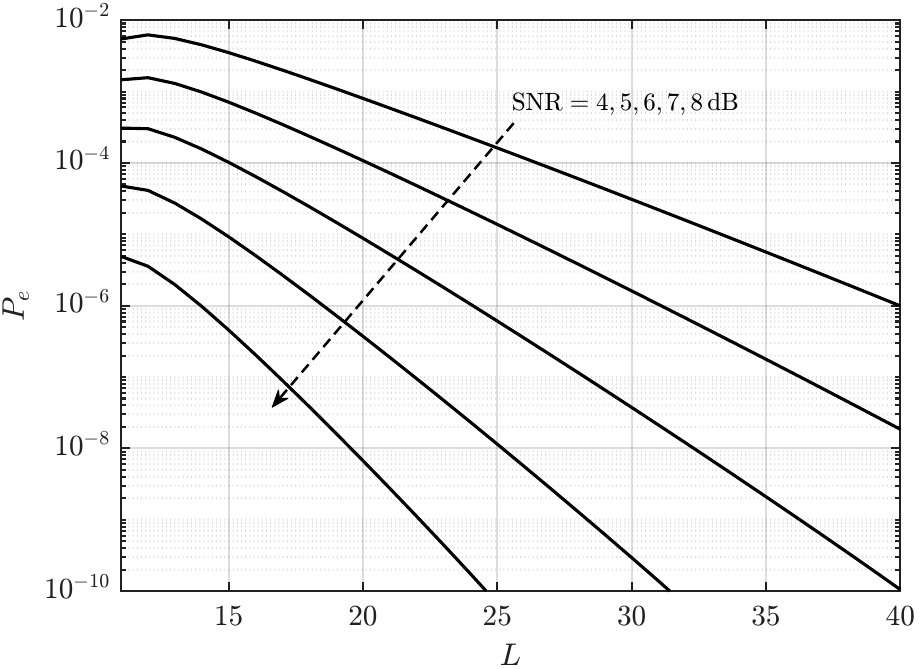}}
    \vspace{-0.2cm}
	\caption{Error probability $P_e$ versus  codeword length $L$ for different SNRs.}
	\label{Fig_04_Pe_vs_L}
\end{figure}
\begin{figure}[!t]
	\centerline{\includegraphics[width=0.9\columnwidth]{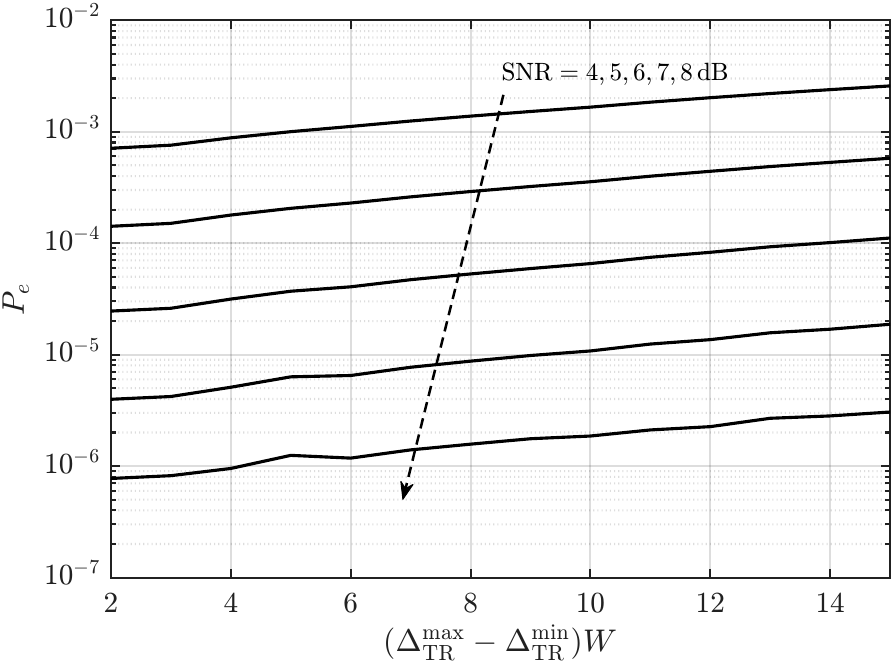}}
    \vspace{-0.2cm}
	\caption{Error probability $P_e$ versus normalized delay spread $(\Delta_{\rm TR}^{\max}-\Delta_{\rm TR}^{\min})W$ for different SNRs.}
	\label{Fig_05_Pe_vs_DelaySpread}
\end{figure}
Fig.~\ref{Fig_03_Rate_vs_L} reports the transmission rate $\mathcal{R}$ versus the codeword length $L$. The rate initially increases rapidly with $L$, attains a maximum at $L=21$ (marked by a filled circle), and then gradually decreases. This behavior arises because the number of transmitted bits grows logarithmically with $L$, while the frame duration increases linearly with $L$, yielding an optimal codeword length that balances these two effects.

Fig.~\ref{Fig_04_Pe_vs_L} reports the error probability $P_e$ versus $L$ for different values of the signal-to-noise ratio (SNR), defined as
\begin{equation}
	\mathrm{SNR}=\frac{\mathcal{P} G B(\bar{\bm{\theta}})\sigma_{\rm TR}^2}{L N \sigma_{{\rm R},\omega}^{2}}.
\end{equation}
Results are averaged over $10^6$ realizations of the TR channel.  For all considered SNR values, $P_e$ decreases as $L$ increases. This behavior is due to the coherent processing gain provided by longer codewords, which enhances the separability between active and inactive codewords in the matched-filter outputs, as also reflected in the first argument of the generalized Marcum $Q$-function in~\eqref{Pc_known_rate}. Increasing the SNR further reduces $P_e$.

Finally, Fig.~\ref{Fig_05_Pe_vs_DelaySpread} reports $P_e$ versus $(\Delta_{\rm TR}^{\max}-\Delta_{\rm TR}^{\min})W$ for different SNR values. It is observed that $P_e$ slightly increases with the normalized delay spread. This is because a larger delay spread widens the observation window: while the signal energy remains essentially constant, the noise accumulates over a larger number of samples $K_{\rm R}$, reducing the separability between codewords at the matched-filter output.

\section{Conclusions}\label{SEC_Conclusions}
This paper proposed a non-linear modulation scheme for RIS-based MIMO radar systems that implement backscatter communication. By encoding information in the unordered selection of orthogonal slow-time codewords, the proposed approach preserves the RIS beampattern and enables low-complexity noncoherent detection based on matched filtering. The resulting rate--reliability tradeoff was also discussed. Ongoing work is investigating encoding and decoding strategies for backscatter communication that enable variable transmission rates within a fixed frame interval. In addition, we are developing processing techniques to operate the radar receiver without prior knowledge of the embedded message.

\balance


\begin{thebibliography}{10}
	\providecommand{\url}[1]{#1}
	\csname url@samestyle\endcsname
	\providecommand{\newblock}{\relax}
	\providecommand{\bibinfo}[2]{#2}
	\providecommand{\BIBentrySTDinterwordspacing}{\spaceskip=0pt\relax}
	\providecommand{\BIBentryALTinterwordstretchfactor}{4}
	\providecommand{\BIBentryALTinterwordspacing}{\spaceskip=\fontdimen2\font plus
		\BIBentryALTinterwordstretchfactor\fontdimen3\font minus
		\fontdimen4\font\relax}
	\providecommand{\BIBforeignlanguage}[2]{{%
			\expandafter\ifx\csname l@#1\endcsname\relax
			\typeout{** WARNING: IEEEtran.bst: No hyphenation pattern has been}%
			\typeout{** loaded for the language `#1'. Using the pattern for}%
			\typeout{** the default language instead.}%
			\else
			\language=\csname l@#1\endcsname
			\fi
			#2}}
	\providecommand{\BIBdecl}{\relax}
	\BIBdecl
	\renewcommand{\BIBentryALTinterwordstretchfactor}{4}
	
	\bibitem{Huynh2018}
	N.~V. Huynh \emph{et~al.}, ``Ambient backscatter communications: A contemporary
	survey,'' \emph{IEEE Communications Surveys \& Tutorials}, vol.~20, no.~4,
	pp. 2889--2922, Fourthquarter 2018.
	
	\bibitem{Stockman-1948}
	H.~Stockman, ``Communication by means of reflected power,'' \emph{Proceedings
		of the IRE}, vol.~36, no.~10, pp. 1196--1204, Oct. 1948.
	
	\bibitem{nikitin2008antennas}
	P.~V. Nikitin and K.~V.~S. Rao, ``Antennas and propagation in {UHF RFID}
	systems,'' in \emph{Proc. IEEE Int. Conf. RFID}, Las Vegas, NV, USA, Apr.
	2008, pp. 277--288.
	
	\bibitem{Liu13ambientbackscatter}
	V.~Liu \emph{et~al.}, ``Ambient backscatter: Wireless communication out of thin
	air,'' in \emph{Proc. of ACM SIGCOMM}, Hong Kong, China, Aug. 2013, p.
	39–50.
	
	\bibitem{BackCom-RIS-2022}
	Y.-C. Liang \emph{et~al.}, ``Backscatter communication assisted by
	reconfigurable intelligent surfaces,'' \emph{Proceedings of the IEEE}, vol.
	110, no.~9, pp. 1339--1357, Sep. 2022.
	
	\bibitem{Liu_RIS_ComSurTut2021}
	Y.~Liu \emph{et~al.}, ``Reconfigurable intelligent surfaces: Principles and
	opportunities,'' \emph{IEEE Communications Surveys \& Tutorials}, vol.~23,
	no.~3, pp. 1546--1577, thirdquarter 2021.
	
	\bibitem{Mizmizi_STcodedRIS_JSAC2024}
	M.~Mizmizi, D.~Tagliaferri, and U.~Spagnolini, ``Wireless communications with
	space–time modulated metasurfaces,'' \emph{IEEE Journal on Selected Areas
		in Communications}, vol.~42, no.~6, pp. 1534--1548, Jun. 2024.
	
	\bibitem{Chen_STcodedRIS_Nature2025}
	X.~Q. Chen \emph{et~al.}, ``Integrated sensing and communication based on
	space-time-coding metasurfaces,'' \emph{Nature Communications}, vol.~16,
	no.~1, p. 1836, 2025.
	
	\bibitem{Basar_Com-RIS_Access2019}
	E.~Basar \emph{et~al.}, ``Wireless communications through reconfigurable
	intelligent surfaces,'' \emph{IEEE Access}, vol.~7, pp. 116\,753--116\,773,
	2019.
	
	\bibitem{Yu_RIS_ISAC_TCOM2023}
	Z.~Yu, X.~Wang, X.~Mu, Y.~Liu, and R.~Schober, ``Active {RIS}-aided {ISAC}
	systems: Beamforming design and performance analysis,'' \emph{IEEE
		Transactions on Communications}, vol.~72, no.~3, pp. 1578--1595, Mar. 2023.
	
	\bibitem{Aubry-2021}
	A.~Aubry, A.~D. Maio, and M.~Rosamilia, ``Reconfigurable intelligent surfaces
	for {N-LOS} radar surveillance,'' \emph{IEEE Transactions on Vehicular
		Technology}, vol.~70, no.~10, pp. 10\,735--10\,749, Oct. 2021.
	
	\bibitem{Buzzi_RIS_SP2022}
	S.~Buzzi, E.~Grossi, M.~Lops, and L.~Venturino, ``Foundations of {MIMO} radar
	detection aided by reconfigurable intelligent surfaces,'' \emph{IEEE
		Transactions on Signal Processing}, vol.~70, pp. 1749--1763, 2022.
	
	\bibitem{Chepuri_RIS-ISAC_SPM2023}
	S.~P. Chepuri \emph{et~al.}, ``Integrated sensing and communications with
	reconfigurable intelligent surfaces: From signal modeling to processing,''
	\emph{IEEE Signal Processing Magazine}, vol.~40, no.~6, pp. 41--62, Sep.
	2023.
	
	\bibitem{Taremizadeh_STAR-RIS_OJCS2025}
	H.~Taremizadeh, E.~Grossi, L.~Venturino, and M.~Lops, ``{ISAC STAR-RIS}
	transceivers with space-time coded pulsed signals,'' \emph{IEEE Open Journal
		of the Communications Society}, vol.~6, pp. 9569--9586, 2025.
	
	\bibitem{Wang-2023}
	X.~Wang, Z.~Fei, and Q.~Wu, ``Integrated sensing and communication for
	{RIS}-assisted backscatter systems,'' \emph{IEEE Internet of Things Journal},
	vol.~10, no.~15, pp. 13\,716--13\,726, Aug. 2023.
	
	\bibitem{Deng_RIS-vehicular_TIV2025}
	M.~Deng \emph{et~al.}, ``Reconfigurable intelligent surfaces enabled vehicular
	communications: A comprehensive survey of recent advances and future
	challenges,'' \emph{IEEE Transactions on Intelligent Vehicles}, vol.~10,
	no.~8, pp. 4191--4216, Aug 2025.
	
	\bibitem{Liang_Symbiotic-Radio_TCCN2020}
	Y.-C. Liang, Q.~Zhang, E.~G. Larsson, and G.~Y. Li, ``Symbiotic radio:
	Cognitive backscattering communications for future wireless networks,''
	\emph{IEEE Transactions on Cognitive Communications and Networking}, vol.~6,
	no.~4, pp. 1242--1255, Dec 2020.
	
	\bibitem{Ren_Symbiotic-Radio_ITJ2023}
	C.~Ren and L.~Liu, ``Toward full passive internet of things: Symbiotic
	localization and ambient backscatter communication,'' \emph{IEEE Internet of
		Things Journal}, vol.~10, no.~22, pp. 19\,495--19\,506, Nov 2023.
	
	\bibitem{Venturino_RadBackCom_TWC2023}
	L.~Venturino, E.~Grossi, M.~Lops, J.~Johnston, and X.~Wang, ``Radar-enabled
	ambient backscatter communications,'' \emph{IEEE Transactions on Wireless
		Communications}, vol.~22, no.~12, pp. 8666--8680, Dec. 2023.
	
	\bibitem{Venturino_RadBackCom_TWC2024}
	L.~Venturino, E.~Grossi, J.~Johnston, M.~Lops, and X.~Wang, ``Semi-blind
	multi-tag ambient backscatter communications using radar signals,''
	\emph{IEEE Transactions on Wireless Communications}, vol.~23, no.~12, pp.
	19\,870--19\,884, Dec. 2024.
	
	\bibitem{LiStoica2007}
	J.~Li and P.~Stoica, ``{MIMO} radar with colocated antennas,'' \emph{IEEE
		Signal Processing Magazine}, vol.~24, no.~5, pp. 106--114, 2007.
	
	\bibitem{Book_StoicaLi_2009}
	J.~Li and P.~Stoica, \emph{MIMO Radar Signal Processing}.\hskip 1em plus 0.5em
	minus 0.4em\relax New York, NY, USA: John Wiley Sons, 2009.
	
	\bibitem{Book_Stutzman_2012}
	W.~L. Stutzman and G.~A. Thiele, \emph{Antenna Theory and Design},
	3rd~ed.\hskip 1em plus 0.5em minus 0.4em\relax New York, NY, USA: John Wiley
	Sons, 2012.
	
	\bibitem{Delia_RIS_TVT2025}
	C.~D'Elia, E.~Grossi, and L.~Venturino, ``Beampattern design for transmit
	architectures based on reconfigurable intelligent surfaces,'' \emph{IEEE
		Transactions on Vehicular Technology}, vol.~74, no.~12, pp. 19\,323--19\,338,
	Dec. 2025.
	
	\bibitem{Qian-2017a}
	J.~Qian, F.~Gao, G.~Wang, S.~Jin, and H.~Zhu, ``Noncoherent detection for
	ambient backscatter systems,'' \emph{IEEE Transactions on Wireless
		Communications}, vol.~16, no.~3, pp. 1412--1422, Mar. 2017.
	
	\bibitem{Book_Papoulis_2002}
	A.~Papoulis and S.~U. Pillai, \emph{Probability, Random Variables, and
		Stochastic Processes}, 4th~ed.\hskip 1em plus 0.5em minus 0.4em\relax New
	York: McGraw-Hill, 2002.
	
	\bibitem{Shnidman-1999}
	D.~A. Shnidman, ``Generalized radar clutter model,'' \emph{IEEE Transactions on
		Aerospace and Electronic Systems}, vol.~35, no.~3, pp. 857--865, Jul. 1999.
	
\end{thebibliography}
\end{document}